# Pilotage des processus collaboratifs dans les systèmes PLM

## Quels indicateurs pour quelle évaluation des performances

*Soumaya El Kadiri\* — Philippe Pernelle\*\* — Miguel Delattre\* — Abdelaziz Bouras\**

*\* Université de Lyon - Université Lumière Lyon 2 – Laboratoire LIESP*
*IUT Lumière - 160, Bd de l'Université - 69676- BRON Cedex – France*

*{soumaya.el-kadiri}{miguel.delattre}{abdelaziz.bouras}@univ-lyon2.fr*

*\*\* Université de Lyon - Université Claude Bernard Lyon 1 – Laboratoire LIESP*
*69622 Villeurbanne – France*

*philippe.pernelle@iutb.univ-lyon1.fr*

*RÉSUMÉ. Les entreprises qui collaborent dans un processus de développement de produit ont besoin de mettre en oeuvre une gestion efficace des activités collaborative. Malgré la mise en place d'un PLM, les activités collaborative sont loin d'être aussi efficace que l'on pourrait s'y attendre. Cet article propose une analyse des problématiques de la collaboration avec un système PLM. A partir de ces analyses, nous proposons la mise en place d'indicateurs et d'actions sur les processus visant à identifier puis atténuer les freins dans le travail collaboratif.*

ABSTRACT. *Companies that collaborate within the product development processes need to implement an effective management of their collaborative activities. Despite the implementation of a PLM system, the collaborative activities are not efficient as it might be expected. This paper presents an analysis of the problems related to the collaborative work using a PLM system, identified through a survey. From this analysis, we propose an approach for improving collaborative processes within a PLM system, based on monitoring indicators. This approach leads to identify and therefore to mitigate the brakes of the collaborative work.*

*MOTS-CLÉS : PLM, collaboration, freins à la collaboration, Indicateurs de suivi, extension des processus collaboratif.*

**1. Introduction**

Les systèmes PLM sont des systèmes d'information utilisés dans les entreprises industrielles qui souhaitent gérer les données liées aux produits ceci à toutes les étapes du cycle de vie (conception, production, recyclage). Choisir un PLM s'inscrit dans une démarche stratégique qui vise à optimiser le développement des produits. Elle se caractérise par la mise en place d'un ensemble d'outils permettant à l'entreprise d'assurer le pilotage, la coordination et la traçabilité des données, des informations et des activités liées au cycle de vie du produit. Cette approche est valable pour les grands groupes industriels mais aussi pour les entreprises de taille moyenne. Toutefois, les PME/PMI restent réticentes à mettre en oeuvre de tels systèmes. En effet, la mise en place d'un PLM modifie en profondeur les modes d'organisation et de fonctionnement de l'entreprise. La volonté affichée de mieux collaborer n'est pas toujours suivie d'effet perceptible, avec même dans de nombreux cas des effets inverses provoquant des blocages.

L'objectif de cet article est de proposer une démarche d'amélioration de la collaboration (interne et externe) avec un PLM au sein d'une entreprise de taille moyenne.

Dans une première étape, nous présentons le cadre de nos travaux et plus particulièrement le projet ANCAR-PLM[1]. L'objectif de ce projet est de contribuer à l'étude et à la mise en oeuvre de méthodologies d'intégration et de déploiement de solutions, souvent qualifiées de systèmes d'information pour le PLM (Product Lifecycle Management), permettent la gestion de l'ensemble des flux d'information autour du produit/service à toutes les phases de son cycle vie et avec l'ensemble des partenaires du projet. Ces systèmes sont principalement utilisés dans une démarche d'optimisation de produits existants (capitalisation, réduction des délais, etc.) ou de développement de nouveaux produits. Dans ce contexte, nous abordons plus spécifiquement les problématiques de la collaboration des PME/PMI au sein des systèmes PLM.
Dans une seconde étape, nous présentons une démarche dont l'objectif est de limiter les freins liés au travail collaboratif. Cette démarche se décompose en deux actions principales : la première explique la construction d'indicateurs de suivi afin d'analyser et repérer les manques de fluidité ou les points de blocage. La seconde propose une extension des modèles de processus afin d'atténuer les points de blocage précédemment identifiés.

**2. La collaboration au sein d'un PLM**

---

[1] ANalyse, CARactérisation et mises en oeuvre de solutions de gestion de cycles de vie de produits et de services

La notion de collaboration est centrale dans une perspective PLM. Les activités industrielles ne peuvent plus faire l'économie d'une action concertée des différentes parties prenantes. Dans le même temps ces dernières induisent des freins qui nécessitent la mise en place de dispositifs et outils de régulations.

**2.1. Processus collaboratifs**

La collaboration au sein des systèmes PLM est essentiellement gérée par les processus métiers. La notion de processus permet de modéliser la dynamique des informations et de définir de manière cohérente les comportements des différents objets manipulés (Labrousse *et al.*, 2004). Nous parlons de processus collaboratif lorsqu'une des activités / tâches du processus est réalisée par un ou plusieurs acteurs (Berztiss, 1999). Elle est nécessairement réalisée par au moins deux acteurs, et peut rassembler des acteurs internes au projet et/ou à l'entreprise, comme des acteurs externes (Rose, 2004). Les besoins en coordination de ces activités, et par la suite en modélisation, sont importants (Georgakopoulos, 2001).

On distingue deux grandes catégories de processus : *répétitif* ou *unique* (Berthier, 2006). Un *processus répétitif* est amené à être exécuté à de multiples reprises et sa description sous forme de modèle présente un caractère normatif : les activités sont supposées être effectuées conformément à leur description et à leur ordonnancement. Un *processus unique*, dont l'exemple principal est le projet, n'est exécuté qu'une seule fois. Dans ce dernier cas, pourquoi le modéliser ? Ce peut être soit pour des raisons de planification (identification des activités et répartition entre un groupe d'acteurs), soit parce qu'on cherche à représenter une structure générique pouvant servir de support à un outil de collaboration autour d'un processus (WfMC, 1999).

Par ailleurs, on distingue trois grandes *approches de structuration d'un processus* (Vidal *et al.*, 2002). Certains processus peuvent être définis par une structure qui rend complètement compte de l'ordre des activités, alors que pour d'autres, il est difficile ou peu efficace d'imposer tous les liens entre les activités.
La première approche est parfois qualifiée de «*mécaniste*» : le rôle du processus est de définir précisément l'ordre et le contenu des activités à effectuer, pour accroître l'efficience (réduction des moyens) et l'efficacité (meilleure atteinte du but) du travail.
Dans la deuxième approche, dite «*systémique*», on considère que les activités sont des composants réagissant à des événements. Les liens entre activités s'effectuent par les résultats: le résultat d'une activité représente un événement déclencheur pour autre activité. Le déroulement réel d'une instance de processus correspondra à l'un des chemins prévus.
Dans la troisième approche, qualifiée d'« *émergente* » ou de « construit social », on ne souhaite pas établir de chemin, même multiple, entre les activités. Ce n'est qu'*a* *posteriori* que l'on peut éventuellement retracer la séquence des activités. Chaque activité est assortie d'événements pouvant la déclencher, l'interrompre ou modifier son cours. Un événement est soit d'origine externe, soit temporel, soit le résultat de la sollicitation d'un autre acteur. Ce type de représentation correspond à un processus dont le déroulement n'est pas déterminé *a priori*, par exemple un processus unique dans lequel les acteurs possèdent une latitude dans la façon dont ils vont accomplir une activité.

L'analyse des processus métiers repose par ailleurs sur une distinction entre les différents processus. Nous pouvons ainsi dissocier les processus variants de ceux pérennes dans le temps (Debaecker, 2004) :
- *Processus pérennes dans le temps* : il s'agit des invariants des processus métier, c'est-à-dire les regroupements d'activités qui sont suffisamment stables et autonomes quelles que soient les évolutions de l'entreprise. Ces invariants correspondent en fait aux grands processus de l'entreprise ;
- *Processus variants* : ils sont remplacés facilement lors des changements.

Nous pouvons également dissocier les processus selon leur finalité. Nous distinguons ainsi trois types de processus :
- *Processus administratifs* : il s'agit des processus de gestion de l'information et des documents nécessaires pour l'accomplissement des tâches de chaque acteur impliqué. Ils sont simples et stabilisées dans le temps ;
- *Processus métiers* : il s'agit des processus liés aux services que l'organisation propose et dont son efficacité dépend. Ils sont généralement plus complexes que ceux administratifs. Il s'agit des processus liés au produit : conception, études, fabrication, …

### 2.2. Collaboration et environnement industriel

Un des objectifs du PLM pour réduire le cycle des produits est de favoriser la collaboration entre les différents acteurs qui participent aux processus métiers de développement des produits.
Traiter du travail collaboratif et/ou coopératif, nécessite de traiter au préalable
quelques particularités qui lui sont imputées :
- Premièrement, l'ordre d'exécution des activités collaboratives est initialement inconnu ou partiellement connu (processus émergents ou peu structurés), et on doit être en mesure de prendre en compte les exceptions et les changements qui peuvent survenir (la dynamicité est une caractéristique essentielle du travail collaboratif) (Wynen, 2003).
- Deuxièmement, lorsque l'on gère les activités coopératives, le facteur humain est important, il devient nécessaire d'être en mesure de supporter la modification du choix et/ou du nombre des partenaires (coopération entre partenaires multiples) (Génelot, 2001).

- Troisièmement, ces activités gèrent des données qui sont partagées et échangées, et on doit pouvoir supporter cet aspect, sachant qu'une donnée peut être possédée par plusieurs partenaires et que les versions doivent être cohérentes pour chacun des partenaires (Sheth et al., 1996).
- Enfin, il faut que les équilibres entre compétitivité et collaboration, entre transparence et confidentialité et entre autonomie et cohérence globale soient respectés (Barrand, 2006), (Ithnin et al., 2006).

Une limite des systèmes actuels tient à leur insuffisante gestion de la flexibilité nécessaire au travail collaboratif – l'aspect dynamique et l'aspect connaissance partielle du processus à exécuter (Carmes *et al.*, 2005). Une gestion centralisée est acceptable dans le cadre d'un projet intra-entreprise, mais ce modèle atteint assez vite ses limites lorsque que l'on parle de procédés répartis dans plusieurs entreprises.

Les travaux du CSCW[2] proposent un cadre méthodologique pour faciliter le travail collaboratif au sein d'un groupe d'acteurs. Un des aspects traités par le CSCW est l'analyse des interactions entre acteurs qui aboutit à une segmentation des activités collaboratives. Ces travaux ont donné lieu à la distinction classique entre les activités : communication, coordination, collaboration.
Cette segmentation est incomplète dans le cadre d'une activité industrielle conduisant au développement de nouveaux produits. En effet, il convient d'avoir une approche organisationnelle car elle induit des contraintes sur modes de collaboration et sur les outils, et notamment sur les processus.

La prise en compte du contexte organisationnel est un des éléments de différentiation majeurs. En effet, selon la structure industrielle (grands groupes, entreprise réseau, sous-traitance, co-conception, …) les collaborations entre les individus sont soumises à des contraintes différentes. A titre d'exemple, on peut citer :
- Les contraintes hiérarchiques qui classiquement imposent des restrictions dans les modes de collaborations si les personnes ne sont pas sur le même niveau.
- Les contraintes fonctionnelles/procédurales sont établies par le mode de fonctionnement de l'entreprise ou par des contraintes externes. Par exemple, la certification ou une réglementation sectorielle (chimie, agro-alimentaire) imposent des procédures à respecter.
- Les contraintes communautaristes plus diffuses mais toujours présentes dans tous les secteurs et quelle que soit l'activité. Elles induisent des biais en fonction du sentiment d'appartenance à un groupe des acteurs qui collaborent.
- Les contraintes client dépendent du degré d'implication des clients dans l'activité de l'entreprise. Dans le cas de la sous-traitance, le client est très présent et les activités collaboratives seront conditionnées par ce paradoxe :

---

[2] Computer Supported Cooperative Work

satisfaire le client en lui donnant le maximum d'information et ne pas livrer son savoir-faire pour maintenir la pérennité de l'entreprise.

Certaines de ces contraintes « agissent » partiellement mais de façon explicite dans le système PLM. C'est le cas des contraintes hiérarchiques qui impactent la gestion des droits d'accès aux objets et la gestion des modifications ; ainsi que les contraintes fonctionnelles/procédurales qui sont naturellement mises en oeuvre dans les workflows du PLM (Chappelet *et al.*, 2001), (Nurcan, 1996). Les problèmes de collaboration viennent en fait de ces contraintes qui « agissent » de façon implicite ou indirectement dans l'activité collaborative

Notre propos n'est pas d'agir sur ces contraintes mais d'analyser les conséquences de celles-ci et de proposer des actions afin d'en diminuer la portée. Dans ce cadre, le chapitre suivant présente la mise en évidence des freins à la collaboration tel que nous les avons identifiés au travers d'une enquête.

## 3. Mise en évidence des freins à la collaboration

### 3.1. Contexte des travaux

Ces travaux se situent dans le cadre du projet ANCAR-PLM [3], projet transversal du cluster GOSPI. Le projet est découpé en quatre axes de travail :
- Etude de l'intégration globale au SI et pérennisation des connaissances
- Etude des solutions d'interopérabilité et de gestion de configuration
- Evaluation adaptation et Intégration de la collaboration
- Méthodologie de déploiement des solutions PLM

Les deux premiers axes visent à proposer un état de l'art préalable sur les dimensions d'intégration et d'interopérabilité afin de mieux cerner les conditions instrumentales nécessaires au développement d'une collaboration centrée sur des systèmes PLM. Nos travaux se déclinent plus particulièrement sur les axes 3 et 4. Ils proposent notamment une réflexion sur la définition d'indicateurs spécifiques qui vont permettre de mettre en place un contrôle a priori et a posteriori de l'activité collaborative autour de l'idée clef suivante : anticiper les problèmes par la mise en place d'indicateurs de suivi de l'activité.
Afin de caractériser les problèmes dans la mise en œuvre et l'utilisation d'un système PLM, nous avons établi un questionnaire qui a été soumis à des entreprises ayant déjà déployé de tels systèmes. L'analyse des résultats de l'enquête nous a permis d'identifier certains points importants.

---

[3] ANalyse et CARactérisation des solutions

### 3.2. Analyse de l'enquête ANCAR-PLM

L'analyse des réponses au questionnaire nous a permis de mettre en évidence certains freins ou points de blocage concernant les processus collaboratif au sein d'un PLM.

Deux catégories de problèmes peuvent être distingués en première analyse :

*1. Manque de souplesse des processus*

La volonté de maîtriser l'ensemble des processus métiers de l'entreprise conduit les responsables à mettre en place des processus sur un très grand nombre de données géré par le PLM. Cela à notamment pour conséquences :
- d'alourdir le travail des acteurs en multipliant les tâches de validation ;
- d'affaiblir les capacités réactives de l'entreprise face au besoin client ;
- de produire des effets de bord sur les données qui sont de fait multi-impacté (surcroît de notification, etc.) ;
- d'alourdir la maintenance de ces systèmes où les processus sont gérés par les cycles de vie.

*2. Manque de souplesse des structures organisationnelles*

La gestion des accès en lien avec les niveaux organisationnels est trop contraignante et ne permet pas une vision globale des capacités d'accès des utilisateurs. Cela est particulièrement vrai en cas de collaboration entre différents partenaires industriels ou la relation de confiance à une importance vitale (réactivité, partage des coûts et de la valeur, …).

Par ailleurs, afin d'appréhender ces problèmes identifiés, si parallèlement l'on analyse le contexte organisationnel des PME/PMI, l'on constate que (Pol *et al.*, 2007):

*1. d'un point de vue organisationnel*

Du fait des petites structures des PME/PMI les différents acteurs se voient attribués différentes tâches et responsabilités dans un projet donné. Cette combinaison des responsabilités donne lieu à des relations informelles, qui jouent un rôle important dans de telles structures.

*2. d'un point de vue processus*

Au niveau macro, les processus métiers sont caractérisés par une formalisation rigide tandis qu'au niveau micro, ils sont caractérisés par une grande flexibilité. Cela s'explique par le besoin de supporter les relations informelles présentes dans l'entreprise.

Finalement, on constate que les freins à un réel travail collaboratif sont surtout liés au niveau de contrôle mis en place. Trop ou pas assez de contrôle (sur les processus et sur l'organisation) nuisent à la collaboration.

A partir d'une deuxième analyse plus détaillée des résultats de l'enquête nous avons décelé quelques causes du blocage au travail collaboratif, en adoptant le ***modèle cause-effet*** (appelé aussi *théorie d'action*) (Bonnefous *et al.*, 2001). Le modèle causes-effets fonde une structure arborescente : un effet peut avoir plusieurs causes, bien identifiables, selon des niveaux successifs de causalité qui nous éloignent de l'objectif visé mais nous rapprochent des potentiels d'action pertinents (agir sur les causes en amont et non sur des symptômes intermédiaires).

Le tableau 1 synthétise l'ensemble des causes identifiées que nous avons classifié selon quatre classes : problèmes organisationnels, problèmes d'individus, problèmes techniques, et problèmes de coordination. Cette liste n'a pas la prétention d'être exhaustive, mais elle nous permet d'appréhender les freins au travail collaboratif.

A titre illustratif, l'une des causes de blocage identifiée s'apparente à des « *Problèmes d'adaptation des acteurs face aux évolutions - [C1.3]* » causés par « *l'instabilité des modèles de données dans le temps - [C1.3.1]* » ou par « *l'instabilité des modèles de processus dans le temps - [C1.3.2]* », générant ainsi des réticences auprès des utilisateurs du système.

Un autre exemple des freins au travail collaboratif est « *le manque d'autonomie des acteurs - [C2.2]* » qui peut être causé soit par « *un problème de répartition des tâches et des responsabilités - [C2.2.1]* » ou encore par « *une limitation des droits d'accès - [C2.2.2]* » qui à son tour peut être causée par « *le niveau de confiance attribué - [C2.2.2.1]* ».

| |
|---|
| **(C1) Organisationnelle** |
| C1.1: Degré de formalisation des procédures de travail |
|     C1.1.2: Problème de réalisabilité (faisabilité) |
|     C1.1.3: Rigidité / Manque de flexibilité |
| C1.2: Contournement des processus |
| C1.3: Problèmes d'adaptation des acteurs face aux évolutions |
|     C1.3.1: Instabilité des modèles de données dans le temps |
|     C1.3.2: Instabilité des modèles de processus dans le temps |
| C1.4: Complexité de la gestion des droits d'accès |
| **(C2) Hommes/Individus** |
| C2.1: Refus d'utilisation du système |
|     C2.1.1: Incompétences, Manque d'expérience |
|     C2.1.2: Changement d'habitudes |
|     C2.1.3: Culture technique |
|     C2.1.4: Attitudes face au partage des informations |
|     C2.1.5: Capacités de réactivité limitées |
| C2.2: Manque d'autonomie |
|     C2.2.1: Problème de répartition des tâches et des responsabilités |
|     C2.2.2: Limitation des droits d'accès |
|         C2.2.2.1: Niveau de confiance |
| **(C3) Technique/Informatique** |
| C3.1: Réseaux |
|     C3.1.1: Surcharge |
|     C3.1.2: Accès lourds aux réseaux des différents sites |
| C3.2: Bugs répétés du système |
| C3.3: Problèmes d'interfaçage avec les outils existants |
| **(C4) Collaboration/Coordination** |
| C4.1: Problèmes de coordination |
|     C4.1.1: Délais de réponse trop longs |
| C4.2: Problèmes de communication |
|     C4.3.1: Crainte de perte du savoir-faire |
|     C4.3.2: Sentiment d'appartenance, d'exclusion |
| C4.3: Outils de collaboration |

**Tableau 1.** *Causes du blocage au travail collaboratif*

A partir des résultats de l'analyse il s'agit de mettre en évidence par des indicateurs les problèmes et d'en déduire une adaptation équilibrée et continue du niveau de contrôle dans l'objectif de faciliter la collaboration. Le chapitre suivant propose de présenter la démarche d'amélioration élaborée.

**4.   Démarche d'amélioration**

Face aux problèmes de collaboration identifiés et présentés dans les paragraphes précédents, nous proposons une démarche d'amélioration visant à limiter les freins liés au travail collaboratif au sein des systèmes PLM (Figure1). Cette démarche se résume en deux actions principales :

1. Construction d'indicateurs de suivi afin d'être capable d'observer et d'anticiper les problèmes de collaboration.

2. Extension des processus collaboratifs afin de pouvoir mettre en œuvre des actions de déblocage des différents problèmes précédemment observés.

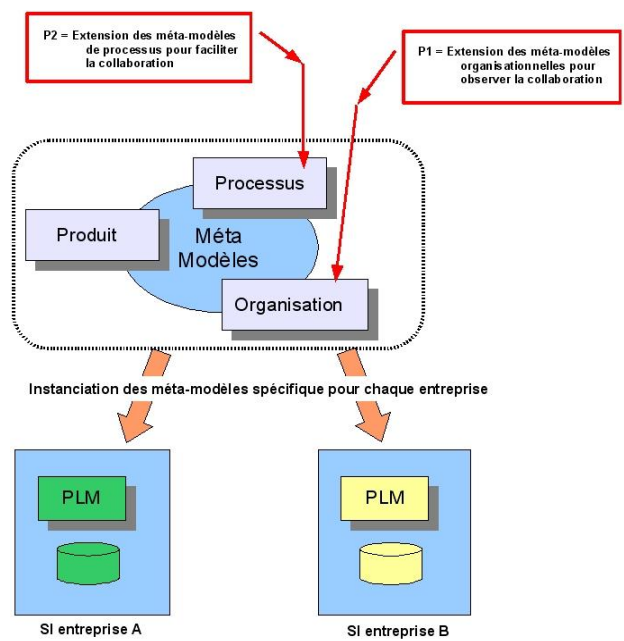

**Figure 1.** *Démarche d'actions sur la collaboration au sein d'un PLM*

Ces actions constituent la mise en œuvre de l'amélioration de la collaboration au sein d'un système PLM. Le paragraphe suivant décrit dans les détails chacune des étapes.

*Action 1 : Construction des indicateurs de suivi*

Nous proposons de construire des indicateurs de suivi afin d'analyser et repérer les manques de fluidité. Leur mise en place permet d'offrir un environnement de gestion de la performance et du pilotage des processus collaboratifs. En effet, la mise en évidence des problèmes rencontrés s'appuie sur l'interprétation d'indicateurs de suivi mis en place. A cet égard, il convient de préciser les différentes dimensions pertinentes liant l'activité collaborative et l'objet manipulé (Mendoza *et al.*, 2002). Celui-ci correspond à l'ensemble des artefacts utilisés au niveau de la collaboration (produit, information, application, service, …). Tout système d'information de type PLM s'appuie sur une structure de méta-modèles (produit, processus, organisation). Tous les éléments des modèles instanciés participent aux développements Produit. Toutefois, tous ces éléments de modèles ne contribuent pas de la même manière aux processus collaboratif. A partir du méta-modèle de données et de processus mis en place, il s'agit d'identifier les objets (classe de document, modèle CAO spécifique, …) et les flux de contrôle présent dans des processus collaboratifs, par le biais des indicateurs.

Afin d'appréhender cette étape de construction des indicateurs de suivi, nous avons adopté le modèle *cause-effet* (Bonnefous *et al.*, 2001). Le principe de ce modèle pour la construction d'un système d'indicateurs est illustré par la Figure 2. **La construction du système d'indicateurs**.

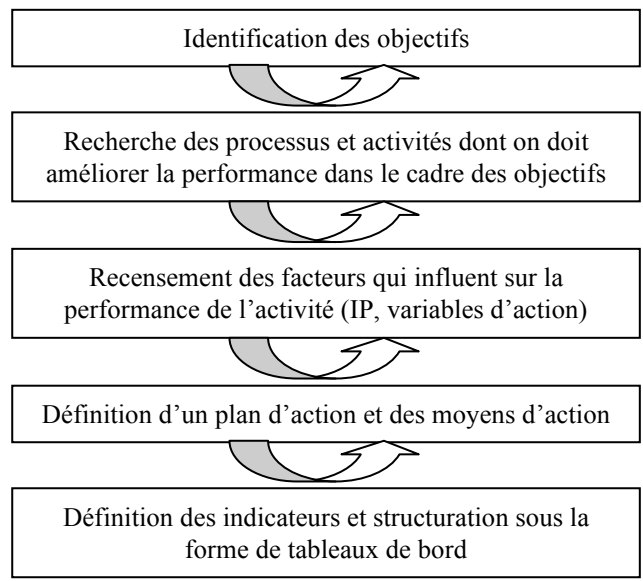

**Figure 2.** *La construction du système d'indicateurs*

Le choix d'un indicateur renvoie nécessairement au choix d'une action comme moyen pertinent d'atteindre un objectif. Le lien de l'indicateur avec l'objectif se fait donc à travers un double processus d'interprétation par les acteurs, résumé par les deux questions suivantes :
- pour atteindre cet objectif, quelle(s) action(s) faut-il engager (interprétation causes-effets)?
- pour évaluer le déroulement ou le résultat de cette action, quelle information faut-il utiliser (interprétation mesure) ?

A ce sujet, l'objectif recherché étant de limiter les causes du blocage. Nous avons donc recensés à partir de l'analyse des résultats de l'enquête (cf. Tableau 1) les facteurs qui peuvent influencer l'atteinte de cet objectif. Suite à quoi, nous avons procédé à l'identification de quelques indicateurs de suivi associés à chacune des causes recensées dans le Tableau 1. Nous présentons dans le tableau suivant quelques exemples.

A titre illustratif, une des causes identifiée est « *l'instabilité des modèles de processus dans le temps* », un des moyens de mesure peut être : « *le nombre de modifications effectuées sur un modèle de processus donné par période* ».

| (C1) Organisationnelle |
|---|
| IP1- C4.1: Nombre de tâches de demandes de modification/validation effectuées sur un même objet |
| IP2- C4.1: Nombre de demandes de modification/validation refusées |
| IP3- C4.3.2: Nombre de modifications effectuées sur un processus |
| IP4- C4.3.1: Nombre de modifications effectuées sur les modèles de données |
| IP5- C4.2: Nombre de fois ou le type d'échange adopté n'a pas été respecté |
| **(C2) Hommes/Individus** |
| IP6- C3.1.5: Temps passé sur une tâche donnée |
| IP7- C3.1.1, C3.1.5: Temps de recherche d'informations sur un même objet |
| IP8- C3.1: Nombre d'utilisateurs |
| **(C3) Technique/Informatique** |
| IP9- C2.1: Nombre de bugs |
| **(C4) Collaboration/Coordination** |
| IP10- C1.1: Délai de réponse aux demandes de modification et validation |
| IP11- C1.1: Nombre de fois où le temps de réalisation d'une tâche n'a pas été respecté |
| IP12- C1.3: Outils de collaboration utilisés |

**Tableau 2.** *Proposition de quelques exemples d'indicateurs de suivi correspondant à la structure causes-effets (Tableau 1)*

Par ailleurs, la qualité des activités collaboratives peut être évaluée de la même manière que les processus logiciels (Mishra *et al.*, 2006), par exemple avec le modèle CMM[4]. L'idée est de modifier et d'enrichir le scénario dans lequel par exemple certaines activités prévues sont systématiquement rajoutées ou supprimées par les utilisateurs. D'où l'intérêt de la deuxième action, qu'est l'extension des processus.

### *Action 2 : Extension des processus collaboratifs*

Les processus collaboratifs sont marqués par leur instabilité et par l'incomplétude et l'incohérence des règles qui les régissent. Le besoin se fait donc sentir de recourir à un processus quasiment permanent d'élaboration de nouvelles règles du jeu visant leur adaptation et leur optimisation. En effet, l'investissement dans un processus de régulation des interactions entre les différents acteurs contribue à maintenir l'équilibre de la coopération. Nous décrivons ainsi le processus de régulation par :
- la ***détection***, basée sur les indicateurs de suivi mis en place, de problèmes ou de points de blocage au niveau du déroulement d'un workflow ;
- l'***adaptation*** basée sur la définition de mécanismes de régulation, visant à proposer une solution au problème détecté ;
- l'***acceptation*** basée sur la concertation de l'ensemble des parties prenantes ; et
- la ***mise en œuvre*** de règles et de protocoles proposés au niveau du modèle de workflow (l'adoption de nouvelles règles).

Ce processus ainsi décrit est illustré par la figure ci-dessous (Figure 3).

---

[4] Capability Maturity Model

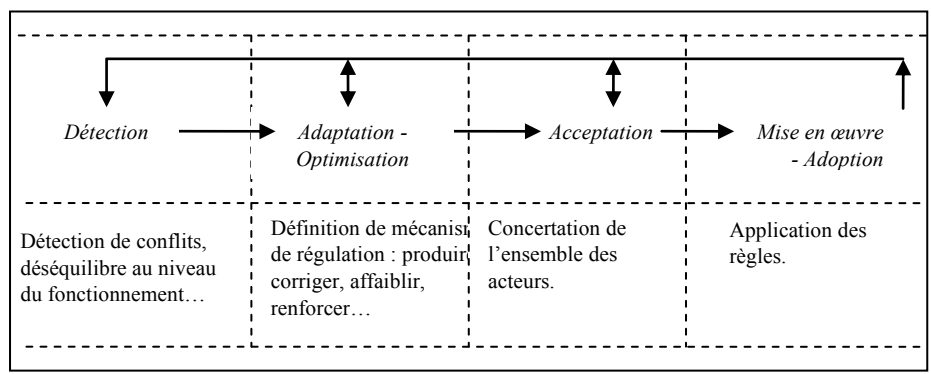

**Figure 3.** *Modèle évolutif d'enrichissement des processus (Régulation)*

Par ailleurs, les différentes expériences dans la mise en place d'un PLM montre qu'il est indispensable de communiquer autour des mécanismes mis en place (Argyris, 1992). En effet, les acteurs doivent disposer d'informations claires et suffisantes sur le déroulement des activités de collaboration, afin d'en limiter les approximations, et les visions partielles (Delattre, 2001). Ils doivent également pouvoir connaître les justifications des choix adoptés, les objectifs, et en percevoir l'impact sur les situations personnelles et collectives. La qualité de la communication découlant de cette démarche méthodologique peut être appréhendé également par des indicateurs de performance, notamment : capacité des utilisateurs à exploiter les ressources potentielles du système, précocité de l'adoption, degré de stabilité (routinisation) de l'utilisation. Ces indicateurs permettront d'évaluer la maturité du projet PLM, et de donner lieu à des initiatives d'amélioration bien ciblées.

L'introduction des concepts d'indicateurs de suivi et de mécanismes de régulations dans la démarche proposée, nous amène à étendre les méta-modèles de processus et organisationnel (Green *et al.*, 2000), (Morley, 2004). Ainsi nous définissons les concepts suivants :
- *Indicateur* : Indicateur de suivi ayant pour objectif de contrôler les activités manipulant les objets du système. Il est formalisé par un objectif, un mode de calcul, et un seuil.
- *Régulateur* : il a pour rôle d'appliquer les mécanismes de régulation et de modération au sein d'une démarche de collaboration. Il peut s'agir d'un tiers ou d'un programme informatique.
- *Règle* : représente les différents mécanismes de régulation mis en place.

La figure ci-dessous illustre l'extension du modèle de classes des processus collaboratifs (Figure 4).

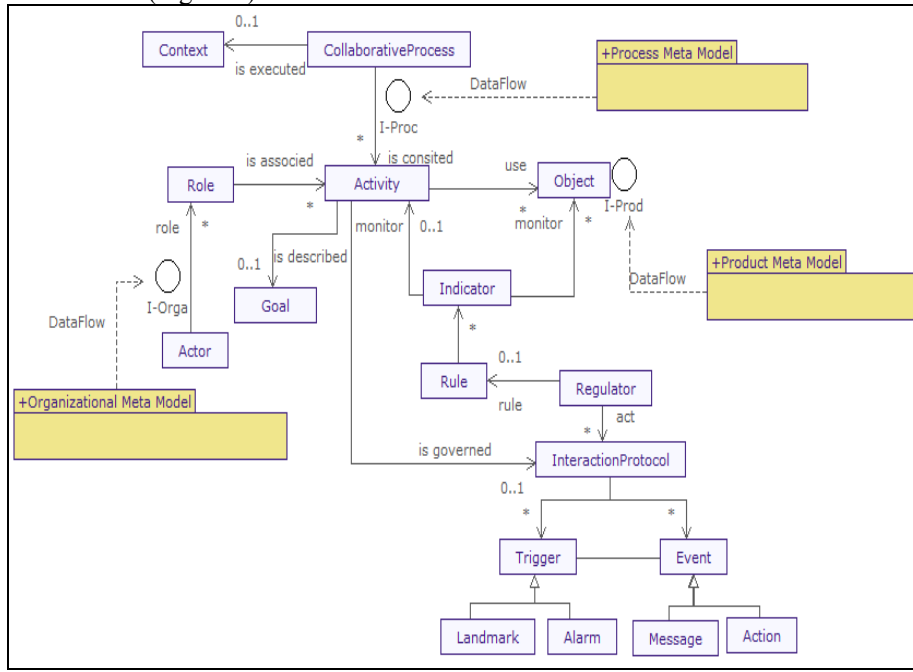

**Figure 4.** *Modèle de classes des processus collaboratifs– Extension des méta-modèles*

## 5. Application

L'application industrielle de nos travaux se situe dans l'environnement de la plasturgie, plus particulièrement dans une PME/PMI du secteur de la transformation de la « La Plastic Vallée» et dont l'activité se positionne sur le secteur automobile (équipementier de deuxième rang). Comme de nombreuses entreprises du secteur, cette entreprise a engagé une démarche d'assurance qualité du type ISO 9001 et ISO TS 16949. Les processus et manuels qualité constituent à ce titre une source importante de connaissance facilitant ainsi la construction des processus métiers. L'activité de l'entreprise étant basé sur des réponses à des appels d'offre, nous avons choisi de concentrer notre application analyse sur ce processus métier. Ce processus implique trois processus interagissant avec différents acteurs internes et externes à l'entreprise. Il se décrit comme suit :

- processus 1 : définition des caractéristiques techniques du produit à développer
- processus 2 : analyse et gestion en interne de l'appel d'offre
- processus 3 : traitement des réponses à l'appel d'offre et prise de décision finale.

L'analyse de l'efficacité de ce processus de collaboration est à ce jour en cours d'expérimentation. Elle sera basée sur les indicateurs proposés dans notre démarche.

## 6. Conclusion

Au terme de ces travaux, nous avons abordé les problématiques de la collaboration des PME/PMI au sein des systèmes PLM. Une étude de terrain des pratiques PLM a été menée afin d'analyser et repérer les manques de fluidité ou les points de blocage au niveau des processus collaboratifs. Une première analyse des résultats de l'enquête nous a permis de déceler deux catégories de problèmes, relativement au travail collaboratif, liés à un manque de souplesse des processus et des structures organisationnelles. Par ailleurs, cette étude a mis en exergue un ensemble de problèmes imbriqués qui s'apparentent à la fois à des contraintes techniques et/ou informatique, à des contraintes organisationnelles, ou encore à des contraintes de coordination/coopération. Cette étude a mis en évidence un réel besoin de méthodologie de mise en place d'indicateurs de suivi.
Cela nous a amené à proposer une démarche méthodologique d'amélioration de la collaboration (interne et externe). Cette démarche consiste en deux actions déterminantes : la première explique la construction d'indicateurs de suivi ; le modèle cause-effet a été adopté pour cette fin. La seconde propose une extension des modèles de processus afin d'atténuer les points de blocage précédemment identifiés.
Par ailleurs, de par la mise en place des indicateurs de mesure de performance, le caractère évolutif de l'approche méthodologique proposée s'inscrit dans une démarche de qualité et d'amélioration continue, permettant d'optimiser les processus en place.
Une suite de ces travaux consiste à compléter l'étape de construction d'indicateurs de suivi. En effet, une critique attribuée à la théorie d'action est que la relation causes-effets peut donner lieu à une structure non linéaire, plus proche d'interactions en boucle, avec des niveaux variables de corrélation entre les facteurs recensés ; et ce davantage lorsqu'on s'éloigne de l'objectif. Ainsi, nous pensons à l'utilisation de la logique floue en vue d'appréhender les corrélations qui peuvent exister entre les différents indicateurs (ou critères de mesure) identifiés, ainsi que d'exploiter le caractère qualitatif qui leur est associé (Berrah, 1997).

La démarche ainsi proposée peut être généralisée et adoptée à tout type de problème autour des systèmes PLM. C'est le cas notamment du problème de capitalisation des connaissances métiers dans les processus de développement de produit.

**Références**